\documentclass[11pt,leqno]{article}
\usepackage{amsmath,amssymb,amsbsy,amsfonts,amsthm,mathtools,mathrsfs}
\usepackage[T1]{fontenc} 			
\usepackage{palatino}				
\usepackage[normalem]{ulem}			
\usepackage{scrextend}				
\usepackage{calc}					
\usepackage{comment}				
\usepackage{enumitem}				
\usepackage{scalerel}				
\usepackage{fontawesome}			
\usepackage{mathabx}				
\usepackage{bbm}					
\usepackage{authblk}				
\usepackage[letterpaper, top=1in, bottom=1in, left=.75in, right=.75in]{geometry}
\usepackage{setspace}				
\usepackage[pdfencoding=auto,psdextra,colorlinks=TRUE,linkcolor=black,citecolor=black,urlcolor=blue]{hyperref}
\usepackage{bookmark}				

\newfont{\boldit}{cmbxti12}
\newcommand{\C}{\mathbbm{C}}  						
\newcommand{\R}{\mathbbm{R}}  						
\newcommand{\N}{\mathbbm{N}}  						
\newcommand{\supp}{\text{supp}}						
\newcommand{\sgn}{\text{sgn}}							
\newcommand\wh[1]{\hstretch{1.111}{\widehat{\hstretch{.9}{#1}}}}	

\newcommand{\DTo}{\mathcal{D}(\R)}						
\newcommand{\DTd}[1]{\mathcal{D}(\R^{#1})}				
\newcommand{\DDo}{\mathcal{D}\,'(\R)}					
\newcommand{\DDd}[1]{\mathcal{D}\,'(\R^{#1})}				
\newcommand{\Lpo}[1]{L^{#1}(\R)}						
\newcommand{\Lpd}[2]{L^{#1}(\R^{#2})}					
\newcommand{\Scd}[1]{\mathcal{S}(\R^{#1})}				
\newcommand{\Wkpd}[3]{W^{#1,\,#2}(\R^{#3})}				
\newcommand{\fall}{\;\;\forall\;}						
\newcommand{\psip}{\psi\,'}							

\renewcommand{\geq}{\vargeq}
\renewcommand{\leq}{\varleq}

\renewcommand{\thefootnote}{\fnsymbol{footnote}}

\makeatletter
\def\thmheadbrackets#1#2#3{%
  \thmname{#1}\thmnumber{\@ifnotempty{#1}{ }\@upn{#2}}%
  \thmnote{ {\the\thm@notefont[\;#3\;]}}}
\makeatother

\newtheoremstyle{brakets}
  {1.5em}
  {1.5em}
  {\itshape}
  {}
  {\bfseries}
  {\\}
  { }
  {\thmheadbrackets{#1}{#2}{#3}}

\theoremstyle{brakets}
\newtheorem{thm}{Theorem}[section]
\newtheorem{cor}[thm]{Corollary}
\newtheorem{defn}[thm]{Definition}

\newtheorem{lem}[thm]{Lemma}

\numberwithin{equation}{section}

\title{The Exact Point Spectrum and Eigenvector of the Unique Continuous $\Lpd{2}{2}$ Bound State Solution to the Dirac Delta Schrodinger Potential in Two Dimensions}
\author{Michael Maroun\footnote{Chief Scientist and Vice President of Research \& Development TeXDyn Industries Corporate Laboratories, Austin, TX}}
\date{\today}

\begin{document}
\maketitle

\renewcommand{\thefootnote}{\arabic{footnote}}

\begin{abstract}
Analyzing the point spectrum, i.e. bound state energy eigenvalue, of the Dirac delta function in two and three dimensions is notoriously difficult without recourse to regularization or renormalization, typically both. The reason for this in two dimensions is two fold; 1) the coupling constant, together with the mass and Planck's constant form an unitless quantity. This causes there to be a missing anomalous length scale. 2) The immediately obvious $L^2$ solution is divergent at the origin, where the Dirac Delta potential has its important point of support as a measure. Due to the uniqueness of the solution presented here, it is immediate that the linear operator (the two dimensional Laplace operator on all of $\R^2$), with the specialized domain constructed here, ensures that the point spectrum has exactly one element. This element is determined precisely, and a natural mathematically rigorous resolution to the anomalous length scale arises. In this work, there is no recourse to renormalization or regularization of any kind.
\end{abstract}

\section{Introduction}

The difficulty of analyzing the attractive Dirac delta stems from the basic fact that the formal symbolic form,
\[
-\,\frac{\hbar^2}{2m}\nabla^2\psi(x) - \alpha\delta(x)\psi(x) = E\psi(x),
\]
for $x\in\R^d$ with ostensibly $\alpha > 0$, $E < 0$ and $\psi(x)\in\Lpd{2}{d}$ does not define a linear operator on this Hilbert space. In fact, it cannot given that $\delta(x)$ is a generalized function and a measure, and for which it lacks a suitable definition of an operator of multiplication\footnote{This is despite the obscure and terse statement made in \cite{Mau}.}. Indeed, the distributional identity $\delta(x)\psi(x) = \delta(x)\psi(0)$ is a serious obstruction. See \cite{ReSi} for more details. In fact, it turns out that delta is compatible with the notion of a quadratic form and it is this quadratic form that can help determine the energy eigenvalue, i.e. point spectrum, which is uniquely associated to the informal expression above, and turns out to be an extension (in the sense of Friedrichs) of the Laplace operator, but with the very particular domain to be constructed.

Despite these seemingly nontrivial technicalities, it turns out for $d=1$ the informal expression above can be solved quite easily. It is worth going through each of the methods to see how each works and how they are different. Nonetheless in $d=1$, they all deliver the same exact result, as any sound method should. What's remarkable is that for $d >1 $ only some of the methods survive as useful. This will be shown explicitly. It is useful to know that throughout this entire work, the relation $A\doteq B$ is to be read as, $A$ is distributionally equivalent to $B$, or $A$ equals $B$ in the sense of distributions.

\subsection{The Method of Integrating the Schrodinger Equation}

This is the method used in \cite{Gri} and it is somewhat surprising that such a straight forward simplistic method could be fruitful given what is known from functional analysis. One wishes to "solve" the quantum system with $x\in\R$,
\begin{equation}
-\,\frac{\hbar^2}{2m}\nabla^2\psi(x) - \alpha\delta(x)\psi(x) = E\psi(x), \label{1}
\end{equation}
for its bound states. On the physical reasoning of the minus sign in front of the potential but with $\alpha > 0$, the highly local potential is regarded as attractive on account of its "graph" lying below the $x$-axis. Thus, its bound states should be in $\Lpo{2}$ and the energy eigenvalue should have $E < 0$. An universal feature is that $psi(x)$ can be constructed everywhere away from the origin, and on this matter the reasoning is solid. Rearrange \eqref{1} omitting the Dirac delta, as shown below, and on account of $d=1$, the formal symbol for the Laplace operator can be replaced by the ordinary second derivative.
\begin{equation}
\psip'(x) - b^2\psi(x) = 0	\label{2}.
\end{equation}
Now this expression is to be solved with the condition that,
\begin{equation}
b = \frac{\sqrt{2m|E|}}{\hbar} > 0, \label{2.1}
\end{equation}
and $\psi(x)\in\Lpo{2}$. One arrives rather quickly at the Hilbert unit normalized expression,
\begin{equation}
\psi(x) =
\begin{cases}
\sqrt{b}\;e^{-b\,x} &,\; x \geq 0 \\
\sqrt{b}\;e^{\phantom{-}b\,x} &,\; x \leq 0,
\end{cases}
\label{3}
\end{equation}
which is not only in $\Lpo{2}$ but in fact it is continuous on all of $\R$ and it is $\Lpo{p} \fall 1 \leq p \leq\infty$. Its derivative however is not continuous, one finds,
\begin{equation}
\psip(x) =
\begin{cases}
-b\,\sqrt{b}\,e^{-b\,x} &,\; x > 0 \\
\phantom{-}b\,\sqrt{b}\,e^{\phantom{-}b\,x} &,\; x < 0.
\end{cases}
\label{4}
\end{equation}
It is easy to see that $\psip(x)$ is also $\Lpo{p} \fall 1\leq p \leq \infty$. But note that it is not continuous at $x=0$. This sole discontinuity motivates the solution by integration. Since $\psi(x)$ has been determined, integrating \eqref{1} should be straightforward. This leads to
\begin{equation}
\begin{split}
-\,\frac{\hbar^2}{2m}\int\limits_{-\infty}^{\infty}\psip'(x)\;dx - \alpha\;\int\limits_{-\infty}^{\infty}\delta(x)\psi(x)\;dx - |E|\int\limits_{-\infty}^{\infty}\psi(x)\;dx =& 0 \\
-\,\frac{\hbar^2}{2m} \psip(x)\big|_{-\infty}^{\infty} - \alpha\psi(0) + |E|\frac{2}{\sqrt{b}} =& 0 \\
0 - \alpha\sqrt{b} + |E|\frac{2}{\sqrt{b}} =& 0. \label{5}
\end{split}
\end{equation}
From the last line, the resulting equality becomes,
\begin{equation}
2|E| = \alpha\;b. \label{6}
\end{equation}
One must now recall that $b$ contains $|E|$ such that $b = \frac{\sqrt{2m|E|}}{\hbar}$. So \eqref{6} now becomes,
\[
\sqrt{|E|} = \frac{\sqrt{m}\alpha}{\sqrt{2}\hbar}.
\]
Thus, one concludes
\begin{equation}
E = -\,\frac{m\alpha^2}{2\hbar^2}. \label{7}
\end{equation}
More will be noted on this method later. Its essential features are two fold. First, the value of $\psi(0)$ exists and is finite and does not result in the energy factoring out of the expression entirely, and second, that $\psip(x)\xrightarrow{x\rightarrow\,\pm\,\infty\;}0$. However, keep in mind that there are an infinity of continuous functions on the Hilbert space which do not tend to zero as $x\rightarrow\pm\infty$. For such an example, see \cite{Gie}.

\subsection{The Distributional Calculus}

This method when handled properly is completely rigorous. Unfortunately in review of literature and peer article submissions, there is an overwhelming tendency to misuse the rules of the distributional calculus nearly as frequently as college freshmen misuse the rules of regular calculus. There is also the unfortunate fact that the method seems not to be universally applicable in answering difficult questions about potentials as generalized functions. The fact that it is insufficient is very telling that there are still missing key mathematical tools in order to handle quantum field theory completely rigorously without recourse to renormalization or regularization.

There are two especially useful features about it. The first is that the quantity $H\psi$ can be regarded precisely as a distribution where the potential is a distribution and the Laplace operator is replaced with the distributional derivative. The vector space property of distributions then assure that there is rigorous meaning to the sum of any two vector elements. In the case at hand, namely, $-\tfrac{\hbar^2}{2m}\nabla^2\psi(x)$ and $\delta\cdot\psi(x)$. The reader is cautioned however that the second distribution may not exist whenever $\psi(\supp(\delta))$ does not exist. In fact, it is the case that sufficient conditions are $\delta\cdot\psi\in\DDd{d}$ if $\psi(\supp(\delta))$ exists and is finite. There are some interesting counter examples that are not so much counter examples but rather demonstrations of the flexibility of equivalence in the sense of distributions. One such known example is to find a distribution $T\in\DDo$ such that $T \doteq \frac{\delta(x)}{x}$, where $\doteq$ means equal in the sense of distributions. One eventually arrives at $T\doteq -\delta\,'$. But there is a price to be paid that is seldom discussed. If $\varphi\in\DTo$ then necessarily $x\varphi\in\DTo$ but it is not true that $\frac{1}{x}\varphi\in\DTo$, despite the fact that it could be true.

There are three central facts to be taken away from this information. The first is that (especially in one dimension) rational functions and tempered distributions play well together. The mathematical fact behind this is the Paley-Wiener theorem for the growth rate of functions and distributions through the Fourier transform. Simply put, the Fourier transform of a tempered distribution is necessarily a function of polynomial (i.e. slow) growth. The second fact is that playing such counter example games in higher dimensions gets extremely tricky almost immediately. One of the many reasons is the loss of the simplicity of the rule of L'H\"{o}pital for indeterminate limits of a single variable. That is, L'H\"{o}pital's rule does not apply to higher dimensions, i.e. more than just one (linearly) independent variable. The third and final fact to take away is the price to be paid in the reduction of the test function space. In the case above, one sees that one cannot admit test functions, $\varphi$, for which $\varphi/x$ is not in $\R$ as $x$ tends to zero. The issue in the present example can be entirely sidestepped by simply regarding the distribution in question as $-\delta\,'$. Then one only needs that $-\varphi\,'(0)$ is finite (real in the case of certain physical applications), and this is assured from the very definition of the test functions, $\varphi\in\DTd{d}$.

Indeed consider $\varphi(x)$ in the Schwartz space of test functions of rapid decay. Then the Fourier transform,
\[
\wh{\:x\varphi\!\left(\!x\!\right)} = \frac{\,1}{-2\pi i}\;\wh{\varphi\,}\,'(k),
\]
where $k\in\R$ is the Fourier conjugate variable, $\widehat{\varphi}$ is also in the Schwartz space by construction, and the prime is the derivative with respect to $k$.

The second useful feature is that the distributional derivative is a bounded linear operator. This is a seemingly shocking statement but its truth is not so hard to see. To show that the distributional derivative is bounded one need only show that it is in fact continuous. Its continuity is derived from the topology the distributions inherit from the test function space.

Recall that the test function space, $\DTd{d}$, is the space of all infinitely differentiable functions with compact support, and that they form a dense subspace of the space of distributions, $\DDd{d}$. One can show that both $\DTd{d}$, and $\DDd{d}$ are in fact Fr\'{e}chet and that there is a sequence of seminorms on the space of distributions, which bounds the absolute value of the Schwart bracket. Sequential continuity follows and boundedness is established by showing that the sequence of seminorms is in fact finite. This is a consequence of not only the algebraic closure of $\DTd{d}$ under differentiation as a vector space but also from the set contracting\footnote{This statement is rather delicate. While it is true that the number of points in the support of the test function's derivatives are decreasing, the collection of which may be a set of Lebesgue measure zero.} fact $\supp\left(\partial^k\varphi\right) \subset \supp(\varphi)$. Hence the summation over $k$ is finite because the sequence generated by the supremum of the $k$-th derivatives terminates.

Finally, the existence of the primitive of a distribution and the implicit pervasive presence of integration by parts is sufficient to establish the operator nature rather than simply an abstract map. See \cite{FrJo} for explicit details discussed in sections 1.3, 2.1, and the appendix. There is a somewhat far reaching large impact consequence of this. It implies that all the literature on the mathematical theory of bounded linear operators and their algebraic structure as applied to quantum field theory might not be completely useless, when a suitable notion of bounded is furnished as elucidated here above. Otherwise, the harmonic oscillator and the unbounded spectrum of free photons would render any traditional notion of a bounded linear operator completely useless.

One more fact that is worth noting is the existence of the Gel'fand triple,
\[
\DTd{d}\subset\Lpd{2}{d}\subset\DDd{d}.
\]
With this in mind, elements of the Sobolev space, $\Wkpd{k}{2}{d}$, can on occasion be treated as proxies for test functions whenever $k\geq 2$, the order of the differential operator ($k=2$ for the Laplace operator). The necessary and sufficient conditions for this to be consistent is iff the statement obtained is unchanged $\forall\;\phi\in\DTd{d}$. However, there are such generalized potentials where this is not possible to attain for all test functions but for which meaningful statements can be made for the Sobolev proxies-- typically the wave function itself.

One proceeds as follows. Define the distribution $H\psi$ by first solving the free differential equation and then applying the distributional Laplace operator. From \eqref{3} the unit normalized Hilbert solution is already known to be,
\[
\psi(x) \doteq \sqrt{b}\,e^{-b\,|x\,|\,}.
\]
Now instead, one calculates its distributional derivatives. The first distributional derivative reads,
\[
\psip(x) \doteq -b\,\sqrt{b}\;\sgn(x)\,e^{-b\,|x\,|\,},
\]
while the second becomes,
\begin{equation}
\begin{split}
\psip'(x) &\doteq b^2\,\sqrt{b}\,e^{-b\,|x\,|\,} - 2b\,\sqrt{b}\;\delta(x)\,e^{-b\,|x\,|\,} \\
&\doteq b^2\,\sqrt{b}\,e^{-b\,|x\,|\,} - 2b\,\sqrt{b}\;\delta(x).
\end{split} \label{8}
\end{equation}
The relation symbol $\doteq$ has been used to indicate equality (only) in the sense of distributions. The distribution $H\psi$ can now be assembled as,
\begin{equation}
\begin{split}
H\psi(x) &\doteq -\,\frac{\hbar^2}{2m}\left[b^2\,\sqrt{b}\,e^{-b\,|x\,|\,} - 2b\,\sqrt{b}\;\delta(x)\right] - \alpha\delta(x)\psi(x) \\
&\doteq -|E|\psi(x) + \frac{\hbar^2\,b}{m}\sqrt{b}\;\delta(x) - \sqrt{b}\,\alpha\,\delta(x).
\end{split}
\label{9}
\end{equation}

To obtain a statement of interest one now utilizes the Schwartz bracket, $\langle\cdot,\cdot\rangle$, which pairs a distribution on the left with a test function on the right. The statement of current interest is whether,
\begin{equation}
\langle H\psi, \varphi\rangle = -|E|\langle\psi,\varphi\rangle, \label{10}
\end{equation}
contains useful information. From the last line of \eqref{9}, and the would-be eigenvalue-like equation in \eqref{10}, one obtains,
\[
-|E|\langle\psi,\varphi\rangle + \frac{\hbar^2\,b}{m}\sqrt{b}\;\langle\delta(x),\varphi\rangle - \sqrt{b}\,\alpha\,\langle\delta(x),\varphi\rangle = -|E|\langle\psi,\varphi\rangle.
\]
Notice that the energy eigenvalue terms of the standard Schrodinger equation cancel! Nonetheless, the above expression yields the identity,
\begin{equation}
\begin{split}
\frac{\hbar^2\,b}{m}\langle\delta,\varphi\rangle -\alpha\langle\delta,\varphi\rangle &= 0 \\
\left[\frac{\hbar^2\,b}{m} - \alpha\right]\langle\delta,\varphi\rangle &= 0.
\end{split}
\label{11}
\end{equation}
In order that the last line of \eqref{11} be an exact equality (not in the sense of distributions) it must be true $\forall\;\varphi\in\DTo$, and therefore it must be the case that only the expression in brackets can vanish. Thus,
\[
\frac{\hbar^2\,b}{m} - \alpha = 0,
\]
and consequently,
\[
b = \frac{m\,\alpha}{\hbar^2}.
\]
Again, recall that $b^2 = 2m|E|/\hbar^2$, and hence,
\begin{equation}
E = - \frac{m\alpha^2}{2\hbar^2}.
\end{equation}

The above result is in precise agreement with \eqref{7} from the previous method of integrating the Schrodinger equation. Note that both methods have the weakness that they rely heavily upon the fact that $\psi(\supp(\delta))\in\R$, and also conveniently of the correct sign because it is mandatory that, logically speaking, one must insist upon $\hbar,\,m,\,\alpha,\,|E| > 0$. It is only then that $E<0$, due to the choice of minus sign in front of the Dirac delta function as a physical bound state potential.

\subsection{The Quadratic Form}

The quadratic form that is constructed must not only take into account the Dirac delta potential but also the distributional derivatives of the wave function. At first, it may seem surprising but this is not at all surprising given that one starts with constructing the wave function $\psi(x)$ in the absence of any potential at all. Using $(\cdot,\cdot)$ as the inner product on the Hilbert space, regard that,
\[
\begin{split}
(H\psi,\,\psi) = -\,\frac{\hbar^2}{2m}\int\limits_{x\in\R}\psi(x)\cdot\psip'(x)\;dx - \alpha\int\limits_{x\in\R}\delta(x)|\psi(x)|^2\;dx &= (E\psi,\psi) \\
&= -|E| \\
&=-\,\frac{\hbar^2}{2m}b^2\int\limits_{x\in\R}|\psi(x)|^2\;dx - \alpha|\psi(0)|^2 \\
&= -|E| - \alpha|\psi(0)|^2 \\
&= -|E| - \alpha\cdot b
\end{split}
\]
The last line above implies either $\alpha=0$ which is consistent with \eqref{2}, i.e. the fact that the system was solved away from the presence of the potential, or that $b=0$, which since both $m$, and $\hbar$ are positive real, it implies $E=0$ and hence $\psi=0$, the trivial solution. Although both possibilities are valid logically and mathematically, they are true but not useful-- the bane of every mathematician everywhere\footnote{In the words of Herman Minkowski, {\it "Nobody has ever noticed a place except at a time, or a time except at a place."}-- from Minkowski's {\it Space and Time} as it appeared in the 1952 Dover reprint of the collected works originally from the Methuen \& Co. Ltd. 1923 series entitled, {\it The Principle of Relativity} (for both publishers), translated by W. Perrett and G. B. Jeffery.}.

This makes it clear that for such highly local singular potentials, the inner product with the usual derivative is insufficient. Instead, one must construct an abstract quadratic form, one that is unlikely to be associated to a linear operator \cite{ReSi}. In every regard, it is nearly the same as the Schwartz bracket but one does not require that the statement hold for all test functions. This seems strange but it turns out that the would-be domain of such a quadratic form may lie outside the space of test functions. Thus, such a requirement may not only be impossible mathematically; but it may be absurd physically on grounds of the inability to connect the statement with an observable state.

The quadratic form is constructed exactly as on the left hand side of \eqref{10}, and results in a very similar statement as \eqref{11} but with $\langle\delta,\varphi\rangle$ replaced with,
\[
\langle\delta,|\psi(x)|^2\rangle = |\psi(0)|^2 = b.
\]
In any case, the above still factors out of the expression and returns the same expected result as in \eqref{11}, namely,
\[
\frac{\hbar^2 b}{m} - \alpha = 0.
\]
One then also deduces correctly that
\[
E = -\,\frac{m\alpha^2}{2\hbar^2}.
\]

The advantage of the quadratic form is that it allows one to arrive at a solution without needing a statement that is true for all test functions. A statement unlikely to be possible in general because it is conjectured in \cite{Mar1} that both singularities and symmetries pair down the test function space expanding the space of distributions to accommodate the variety of quantum fields observed in nature. Specifically, it allows one to reference a state as perhaps a Sobolev proxy in lieu of a test function, especially where no notion of unitary equivalence is appropriate because neither the Schwartz space nor its continuous topological dual are Hilbert spaces\footnote{As Von Neumann said to Schwinger in a printed recollection of A. Wightman, "On a Hilbert space, a unitary operator always has an inverse.". So too here is it stated, a Hilbert space is always complete with respect to its inner product. The Schwartz function space is not (with respect to the same inner product).}. Through this manner alone, the difficulties of the Haag theorem are evaded. This is beside the observed fact that in nature the Wightman Axioms are inappropriate on physical grounds that table top condensed matter experiments test and observe quantum field theories that are not manifestly relativistically covariant nor Lorentz invariant, unitary equivalence aside.

\subsection{The Resolvent}

The major advantage of using the resolvent operator is that it directly incorporates the highly local singular potential, the Dirac delta function. The constructed solutions away from the origin, while sounding plausible, were more of an heuristic argument of what ought to be true. In the resolvent method it is shown that the poles of the time independent Green's function, indeed do correspond to the same bound state energy. The greatest disadvantage is that it fails exactly when one needs it to succeed. In other words for difficult to analyze systems like the derivative of the Dirac delta function in one dimension, or the current system of the Dirac delta in two dimensions, the resolvent method merely verifies that there is a missing length scale and that the coupling constant, $\hbar$, and the mass $m$ combine (typically in products of integer powers) to form an unitless quantity.

Of course, it is always possible to furnish the system with extra external information to remove the ambiguities. In this work, it is the continuity of the wave function, which is unique and rather natural. While in other works in the references, it is a set of somewhat abstract relations some times of Dirichlet type, some times of Neumann type, and some times of mixed type. All that is known at present is that they produce an answer that has been arrived at by some interesting potentially rigorous methods like nonstandard analysis (in the sense of A. Robinson), and other more common techniques such as those used in theoretical physics, like renormalization and regularization. Incidentally, there are two distinct notions of regularization, one in the physics literature, and one in the mathematics literature. In the mathematics literature on distributions a precise definition is given in \cite{AlG} 3.4, for example. But the word persists in many other subfields of mathematics and there are numerous definitions some more general some less general but they revolve around the same basic notions. The regularization in physics lacks a rigorous definition typically. Some times, it is zeta regularization, some times it is the regularization mentioned in the reference. The set of physicist regularizations and the set of mathematician regularizations have non-empty intersection, but they are far from the same. Which set is larger? Is one fully contained in the other? The answers to these questions seem to be a matter of historical record, both a function of the accuracy of historical record, and a function of what the future to come will bring.

For a linear operator of interest acting on a function, in the present case $(H+\mathbbm{1}|E|)\psi(x)=0$, the resolvent operator is the inverse operator $(H+\mathbbm{1}|E|)^{-1}$. The Fourier transform is of central usefulness here because it is an isometric automorphism on the Hilbert space\footnote{Actually, this is the case by construction for the Schwartz space of test functions. However by the Hahn-Banach theorem, it can be uniquely extended to the Hilbert space of square integrable functions, quite generally so that $f:\C\to\C$ while also, $|f|^2\in\Lpo{1}$.}. The other reason is that the Dirac delta function is the convolution identity, the Fourier convolution theorem can render many differential operators rather easy to manipulate. Still, the reason for bringing up this technical detail is that more often than not, a Green function may be a distribution and not an element of the Hilbert space. This does not invalidate the Green function, rather one is then obliged to check the validity of the statement in the sense of distributions. One commonly denotes this resolvent as $G(x)$ and it will, by construction, necessarily have the property that $(H+\mathbbm{1}|E|)G=\delta$. The original one dimensional system being,
\[
-\frac{\hbar^2}{2m}\psip'(x) - \alpha\delta(x)\psi(x) + |E|\psi(x) = 0
\]
One then wishes to compute $G(x)$ appearing in,
\[
\left[-\frac{\hbar^2}{2m}\frac{d^2}{dx^2} - \alpha\delta(x) + |E|\right]G(x) = \delta(x),
\]
or in a more ready-to-Fourier transform presentation,
\begin{equation}
-\frac{\hbar^2}{2m}G\,''(x) - \alpha\delta(x)G(x) + |E|G(x) = \delta(x) \label{12}
\end{equation}
Apply the Fourier transform to \eqref{12} above, and denote the transform of $G(x)$ as $\wh{G\,}(k)$, where $k$ is the Fourier conjugate variable in the unitary form of the transform with forward kernel $e^{-2\pi i\,k\,x}$. One then obtains the purely algebraic expression,
\[
\frac{4\pi^2\hbar^2}{2m}k^2\wh{G\,}(k) - \alpha G(0) + |E|\wh{G\,}(k) = 1,
\]
from which can be deduced,
\begin{equation}
\wh{G\,}(k) = \frac{1 + \alpha G(0)}{\frac{4\pi^2\hbar^2}{2m}k^2 + |E|}. \label{13}
\end{equation}
There are three important facts to note about \eqref{13}. First, it is an element of $\Lpo{2}$ for $k\in\R$, hence the Fourier transform and its inverse are completely justified. Second, there is a self-referential constant term in the numerator $G(0)$ that is as yet undetermined until the inverse Fourier transform is applied, and a self-consistency check is made. Clearly, the resolvent method in this form may fail if $G(0)\notin\C$. The third and final observation is that $\wh{G\,}(k)$ is even, and so its integral against the imaginary odd component of the Fourier kernel is zero. Thus, the relevant integral identity is,
\[
\int\limits_{k\in\R} \frac{\cos(2\pi k x)}{k^2 + a^2}\;dk = \frac{\pi}{a}\,e^{-2\pi\,a\,|x\,|}
\]
for $a,\,x\in\R$. Notice the sign of the constant $a$ may effect the properties of the answer. But in this case, $a = \frac{\sqrt{2m|E|}}{2\pi\hbar} = b/2\pi > 0$ in accordance with condition \eqref{2.1}.

After proper identification of the constants, one arrives at the intermediate statement,
\begin{equation}
G(x) = \frac{m}{b\,\hbar^2}\left[1 + \alpha G(0)\right]\,e^{-b\,|x\,|}, \label{14}
\end{equation}
where the constant $G(0)$ still must be determined. Evaluating the above expression at $x=0$ attains this and one finds,
\begin{equation}
G(0) = \frac{m}{b\,\hbar^2-\alpha\,m}. \label{15}
\end{equation}
Substituting \eqref{15} into \eqref{14} gives,
\begin{equation}
G(x) = \left[\frac{m}{b\,\hbar^2-\alpha\,m}\right]\,e^{-b\,|x\,|}. \label{16}
\end{equation}
Now the beautiful aspect of the resolvent method (when it is applicable) can be seen. The function $G(x)$ is not only in $\Lpo{2}$, as it must be because it was the Fourier transform of another element of the same Hilbert space, but the pole of the function, i.e. when the denominator of $G(x)$ is zero occurs exactly when,
\[
b = \frac{m\alpha}{\hbar^2}.
\]
This leads to exactly the same and correct bound state energy of
\[
E = -\,\frac{m\,\alpha^2}{2\hbar^2}.
\]

\section{The Unique Continuous Solution for the 2D Schrodinger Equation with a Delta Potential}

Abstractly, one is interested in the solution to the ill-posed expression,
\[
-\,\frac{\hbar^2}{2m}\nabla^2\psi(x) - \alpha\delta(x)\psi(x) = - |E|\psi(x),
\]
for $x\in\R^2$, and $m,\,\hbar,\,\alpha > 0$ and $E < 0$. As before, one takes $b:=\sqrt{2m|E|}/\hbar > 0$. To define the problem rigorously, one first solves the differential equation away from the potential and then applies any of the first three methods discussed in the introduction. The reader is encouraged to investigate the resolvent method to learn why it fails. The answer succinctly is that there is a missing length scale, hence the constants of the system, $\alpha$, $\hbar$, and $m$ form an unitless constant. So there is a pole to the resolvent operator but though it is true; it is unfortunately also unhelpful because it fails to include the parameter $b$, i.e. any reference to the energy of the system\footnote{This is after having to cope with the increase in dimensionality from $d=1$ to $d=2$, while the order of the Schrodinger equation remains fixed at 2, owing to the Laplace operator. Therefore, the application of the Fourier transform becomes more delicate but still straightforward.}.

In addition, the fundamental solution of the Helmholtz equation is singular at the point support of the delta pseudo-potential. For $\R^3$, this is not an obstruction as shown in \cite{Mar2}; but for $\R^2$, it is. To this end, one first defines what it means for the potential to be singular and then must give precise mathematical meaning to the abstract and ill-defined expression $H_o + V(x)$, when $V(x)$ is not an operator of multiplication on the usual Hilbert space of Lebesgue square integrable functions, but rather a distribution.

\begin{defn}[Locally Square Integrable Functions v1] \label{LSIFv1}
Let $T:\R^d\to\C$, then one says $T\in L^2_{\text{loc}}(\R^d)$ if it is the case that
\[
\int\limits_{x\,\in K\subset\R^d} |T|^2\;\text{d}x < \infty \quad\fall K\subset\R^d.
\]
\end{defn}

\begin{defn}[Locally Square Integrable Functions v2] \label{LSIFv2}
\[
 L^2_{\text{loc}}(\R^d) := \left\{\; T\in\DDd{d}\,:\forall\;\text{\raisebox{2pt}{$\varphi$}}\in\DTd{d},\;\text{one has}\;\, (\text{\raisebox{2pt}{$\varphi$}} \,T)\in\Lpd{2}{d} \;\right\}
\]
\end{defn}

\begin{lem}[Bourbaki-Strichartz Equivalence]\label{lemBSE}
\phantom{Test this line!}
	\begin{center}
{\bf {\it Definition\;\ref{LSIFv1}}} $\Longleftrightarrow$ {\bf {\it Definition\;\ref{LSIFv2}}}
	\end{center}
\end{lem}

\begin{proof}[Proof of Lemma \ref{lemBSE} (Bourbaki-Strichartz Equivalence)]
	\noindent Clearly,
	\[
	\|\text{\raisebox{2pt}{$\varphi$}}\|^2_{\text{\raisebox{-2pt}{$\infty$}}}\!\!\!\int\limits_{x\,\in K} |T|^2\;\text{d}x \geq
	\int\limits_{x\,\in K} |\text{\raisebox{2pt}{$\varphi$}}|^2\,|T|^2\;\text{d}x = \int\limits_{x\,\in U} |\text{\raisebox{2pt}{$\varphi$}}\,T|^2\;\text{d}x
	\]
	Let $d := d(K,\,\partial U) > 0$, and $\varepsilon$ be such that $d > 2\varepsilon > 0$. Now consider the compact sets $K_{\varepsilon}$ and $K_{2\varepsilon}$ as closed $(\varepsilon,\,2\varepsilon)$-neighorhoods of $K$. One  has the set inclusion $K\subset K_{\varepsilon}\subset K_{2\varepsilon}\subset U$ with $d_{K_\varepsilon} - \varepsilon > \varepsilon > 0$.	Now define a mollifier, \raisebox{2pt}{$\varphi$}$_\varepsilon$, and take $\text{\raisebox{2pt}{$\chi$}}_{_{K_\varepsilon}}$ as the characteristic function supported on the compact set $K_\varepsilon$. The convolution \raisebox{2pt}{$\chi$}$_{_{K_\varepsilon}}\!\!\!\ast\text{\raisebox{2pt}{$\varphi$}}_{\varepsilon} =:\text{\raisebox{2pt}{$\varphi$}}_{_K}$ together with its inequalities lead to $|\text{\raisebox{2pt}{$\varphi$}}_{_K}|^2 \geq
	|\text{\raisebox{2pt}{$\chi$}}_{_{K}}|^2$. Hence, one sees
	\[
	\infty > \int\limits_{x\,\in U} |T|^2\, |\text{\raisebox{2pt}{$\varphi$}}_{_K}|^2\;\text{d}x \geq
	\int\limits_{x\,\in U} |T|^2\, |\text{\raisebox{2pt}{$\chi$}}_{_K}|^2\;\text{d}x = \int\limits_{x\,\in K} |T|^2\;\text{d}x. \qedhere
	\]
\end{proof}

\begin{defn}[Singular Distribution]
A distribution $T\in\DDd{d}$ is said to be {\bf singular} if it is not regular.
\end{defn}
\begin{defn}[Regular Distribution]
A distribution $T\in\DDd{d}$ is said to be {\bf regular} if $\exists\;f:\R^d\to\R$ s.t.
\[
\int\limits_{x\in\R^d} f(x)\varphi(x)\text{d}\mu(x)\fall\varphi\in\DTd{d}
\]
where d$\mu(x)$ is Lebesgue measure, written simply $dx$ heretofore.
\end{defn}
\begin{defn}[Globally Singular Function]
A function $f:\R^d\to\C$ is {\bf globally singular} if $f\notin\Lpd{p}{d}$ for any $p\geq 1$.
\end{defn}
\begin{defn}[Locally Singular Function]
A function $f:\R^d\to\C$ is {\bf locally singular} if $f\notin\text{L}^p_{\text{loc}}(\R^d)$ for any $p\geq 1$.
\end{defn}
Note that as a result of the above definitions, the 2d Dirac potential is not even $\Lpd{p}{2}$. In addition, one sees how distributions are inherent to keeping a local $\Lpd{2}{d}$ structure, and hence vital to keeping a probability interpretation in the case of both bound and scattering states. The following problem arises regarding the usual inner product. Solving for $\psi$ away from the singular point means, $(H_o-E)\psi = 0$. This then implies that for the inner product, $(\cdot,\,\cdot)$ on $\Lpd{2}{d}$, one gets
\begin{align*}
	(H_\alpha\psi,\,\psi) &= (H_o\psi - \alpha\delta\psi,\,\psi) \\
	&=E - \alpha|\psi(0)|^2 = E.
\end{align*}
This implies either $\alpha=0$ and $E\in(-\infty,\,0]$, or $\psi(0)=0$, which implies the presence of the delta function is not ``detectable'' by $\psi$ making $E$ indeterminate.

\subsection{Generalized Quantum Theory via Distributions}

To remedy this, re-interpret the Laplace operator as the distributional Laplace operator, and replace the $\Lpd{2}{d}$ inner product, $(\cdot,\,\cdot)$ with the Schwartz bracket $\langle\cdot,\,\cdot\rangle$ by working on the Gelfand triple, $\DTd{d}\subset\Lpd{2}{d}\subset\DDd{d}$, where $\DTd{d}$ is the space of smooth test functions with compact support, and $\DDd{d}$ its topological dual.
\begin{defn}[{\bf The C-Spectrum}]
Whenever,
\[
\langle H\psi,\,\text{\raisebox{2pt}{$\varphi$}}\rangle = E\langle\psi,\,\text{\raisebox{2pt}{$\varphi$}}\rangle\fall\text{\raisebox{2pt}{$\varphi$}}\in\DTd{d},
\]
with $(H\psi)\in\DDd{d}$ then $E$ is said to be in the {\bf C-Spectrum} of $H\psi$.
\end{defn}
\noindent To broaden the notion of the {\bf C-Spectrum}, and ensure it includes the usual point spectrum, one adds the next definition.
\begin{defn}[{\bf Proxy Test Functions}]
Whenever,
\[
\langle H\psi,\,\psi\rangle = E\langle\psi,\,\psi\rangle
\]
for $\Lpd{2}{d}\ni\psi\notin\DTd{d}$ then $\psi$ is said to be a {\bf proxy test function}.
\end{defn}
Together with the approximation theorem for distributions stated below as a lemma, there then is a robust formalism that has a precise mathematical structure.
\begin{lem}[{\bf Test Sequence Convergence (Approx Thm)}]
Every proxy test function $\Lpd{2}{d}\ni\psi\notin\DTd{d}$ is necessarily given by some sequence $\text{\raisebox{2pt}{$\varphi$}}_n\in\DTd{d}\fall n\in\N$ such that $\text{\raisebox{2pt}{$\varphi$}}_n \xrightarrow{n\to\infty}\psi$.
\end{lem}
\begin{proof}
$\DTd{d}$ is dense in $\Lpd{2}{d}$, and $\Lpd{2}{d}$ is the Cauchy completion of $\DTd{d}$ in its inner product.
\end{proof}

\subsection{The Unique Continuous Solution}

There are now two classic results that are relevant.
\begin{lem}[Fundamental Solution]
The Fundamental Solution to the Helmholtz equation $-\nabla^2\psi + b^2\psi = \delta$ for $b>0$ and $\psi\in\Lpd{2}{2}$ is $\psi = K_o(b|x|)$, $K_o$ is the MacDonald function of order (sometimes index) 0.
\end{lem}
\begin{proof}
The Fourier transform is unitary on $\Lpd{2}{2}$. The Helmholtz equation turns into an algebraic relation that if inverse transformed, gives the result. \qedhere
\end{proof}
\begin{lem}[Linearly Independent Smooth Solution]
The two linearly independent smooth solutions to the Helmholtz equation $-\nabla^2\psi + b^2\psi = 0$ for $b>0$ is $\psi = [C_1 I_\nu(b|x|) + C_2 K_\nu(b|x|)]e^{i\nu\phi}$, where $K_o$ is the MacDonald function, $\mathrm{I}_o$ the Basset function, and $C_1,C_2\in\R$.
\end{lem}
\begin{proof}
The Laplace operator on $\R^2$ yields the modified Bessel equation of order $\nu$. \qedhere
\end{proof}
The proofs are terse because the results are classical and well-known, typically, they are encountered during graduate education in the mathematical sciences, if not advanced undergraduate courses.
\begin{thm}[Unique Continuous Solution]
For $x\in\R^2$, $E<0$, $m,\,\hbar,\,b=\sqrt{2m|E|}/\hbar > 0$, $\alpha\in\R\setminus\{0\} $, and the distribution $(H\psi)\in\DDd{2}$ given by \vspace{-.5em}
\[
H\psi \,\dot{=}-\frac{\hbar^2}{2m}\Delta\psi(x) -\alpha\,\delta(x)\psi(x) \vspace{-.5em}
\]
\noindent whence $\psi(x)$ is the following unique continuous solution to the equation \vspace{-.5em}
\[
-\frac{\hbar^2}{2m}\nabla^2\psi(x) + |E|\psi(x) = 0 \vspace{-.5em}
\]
\noindent such that $\psi\in\Lpd{2}{2}$, and $\langle\psi,\,\psi\rangle = (\psi,\,\psi) = 1$. One obtains,\vspace{-.75em}
\[
\psi_c(x) = \frac{b}{\sqrt{\pi}u_o\beta}\left[\mathrm{I}_o(b|x|) h(u_o - b|x|) + K_o(b|x|) h(b|x| - u_o)\right]. \vspace{-.75em}
\]
\noindent For $r>0$, $u_o$ is the only solution to $\mathrm{I}_o(r)\!=\!K_o(r)\!$ and one defines $\beta^2 = K_1(u_o)^2 - \mathrm{I}_1(u_o)^2$, for brevity and clarity, and $h(r)$ is the Heaviside function, the primitive of the Dirac delta.
\end{thm}
\begin{proof}
The two solutions $\mathrm{I}_o(b r)$ and $K_o(b r)$ are strictly monotonically increasing (respectively decreasing), and thus have exactly one and only one point of intersection for $r>0$, define it as $u_o$. Cutting the functions off at this point by making use of the Heaviside step function produces a function which is both continuous on all of $\R^2$ owing to the smoothness of $\mathrm{I}_o$ at the origin and square integrable i.e. \vspace{-.5em}
\[
\int\limits_{x\in\R^2} \frac{1}{\pi R^2\beta^2}\left[\mathrm{I}_o(b|x|) h(u_o - b|x|) + K_o(b|x|) h(b|x| - u_o)\right]^2 dx = 1, \vspace{-.5em}
\]
with $u_o \approx 0.4322837\ldots$. In performing the integral, the quantity,
\[
\beta^2 = K_1(u_o)^2 - \mathrm{I}_1(u_o)^2,
\]
arises naturally. \qedhere
\end{proof}
One can now state precisely the results of the generalized quantum theory for this specific system. That is one arrives at the following corollary for the C-Spectrum.
\begin{cor}
For the (family) of states $\psi_c$, the C-spectrum is given by,
\[
E^C = -\frac{m \alpha^2}{2\pi^2 [K_1(u_o) - \mathrm{I}_1(u_o)]^2\,R^2\,\hbar^2}
\]
\end{cor}
\begin{proof}[Proof of the C-Spectrum]
Using the distributional Laplace operator on $\psi_c$ with $\phi\in\Scd{2}$, one gets,
\begin{align*}
\langle H\psi,\,\phi\rangle &= E \langle\psi,\,\phi\rangle \\
&= \langle-\tfrac{\hbar^2}{2m}\Delta\psi - \alpha\delta\psi,\,\phi\rangle \\
&= E \langle\psi,\,\phi\rangle + \tfrac{\hbar^2}{2m}b N\langle[-\mathrm{I}_1(u_o) + K_1(u_o)]\delta(r-R) - \alpha N\delta(x),\,\phi\rangle.
\end{align*}
The previous equation implies
\begin{align*}
\langle -\tfrac{\hbar^2}{2m}b N \mathrm{I}_1(u_o)\delta(r-R) + \tfrac{\hbar^2}{2m}b N K_1(u_o)\delta(r-R) - \alpha N\delta(x),\,\wh{\phi}\rangle &= 0 \\
\langle b \mathrm{I}_1(u_o)\delta(r-R) - b K_1(u_o)\delta(r-R) + \tfrac{2m\alpha}{\hbar^2} \delta(x),\,\wh{\phi}\rangle &=0 \\
\langle 2\pi R b e^{i\theta}[ K_1(u_o) - \mathrm{I}_1(u_o) ] - \tfrac{2m\alpha}{\hbar^2},\,\phi\rangle &=0
\end{align*}
Hence
\[
b = e^{-i\theta}\frac{m\alpha}{\pi R [K_1(u_o) - \mathrm{I}_1(u_o)]\hbar^2}
\]
Since $E^C = -\tfrac{\hbar^2}{2m}|b|^2 < 0$, one gets,
\[
E^C = -\frac{m \alpha^2}{2\pi^2 R^2 [K_1(u_o) - \mathrm{I}_1(u_o)]^2\hbar^2} \qedhere
\]
\end{proof}

\end{document}